\documentclass[prd,reprint,superscriptaddress]{revtex4-1}

\usepackage{amsmath}
\usepackage{relsize}
\usepackage{amssymb}
\usepackage{slashed}
\usepackage{graphicx}
\usepackage{epstopdf}
\usepackage{braket}
\DeclareGraphicsExtensions{.eps}
\DeclareGraphicsExtensions{.pdf}

\usepackage{mathptmx}   

\begin{document}

\title{Lattice corrections to the quark quasidistribution at one-loop}

\author{Carl E. Carlson}\email{carlson@physics.wm.edu}
\affiliation {Physics Department, College of William and Mary, Williamsburg, Virginia 23187, U.S.A.}

\author{Michael Freid}\email{mcfreid@email.wm.edu}
\affiliation {Physics Department, College of William and Mary, Williamsburg, Virginia 23187, U.S.A.}
\affiliation{Thomas Jefferson National Accelerator Facility, Newport News, Virginia 23606}

\vskip 0.5cm
\date{\today}

\begin{abstract}

We calculate radiative corrections to the quark quasidistribution in lattice perturbation theory at one loop to leading orders in the lattice spacing.  We also consider one-loop corrections in continuum Euclidean space.  We find the infrared behavior of the corrections in Euclidean and Minkowski space are different.  We explore features of momentum loop integrals and demonstrate why loop corrections from the lattice perturbation theory and Euclidean continuum do not correspond with their Minkowski brethren, and comment on a recent suggestion for transcending the differences in the results.  Further, we examine the role of the lattice spacing $a$ and of the $r$ parameter in the Wilson action in these radiative corrections.

\end{abstract}

\maketitle

\section{Introduction}

Hadronic structure remains one of the open questions in nuclear physics.  This structure is governed by Quantum Chromodynamics (QCD).  However, QCD is fundamentally non-perturbative at low energies and thus the interactions among the quarks that make up each hadron are not easily calculable from first principles.  

Deep inelastic scattering (DIS) has been successful at investigating features of QCD such as the distribution of quarks within a hadron~\cite{Alekhin:2012ig,Gao:2013xoa,CooperSarkar:2011aa,Martin2009,Ball:2012cx,Radescu:2010zz,Jimenez-Delgado:2014xza,Jimenez-Delgado:2013boa,Ball:2013lla,Arbabifar:2013tma,Owens:2012bv,Accardi:2011fa,Leader:2010rb,Blumlein:2010rn,MoosaviNejad:2016ebo,Khanpour:2016pph,Shahri:2016uzl}.  Typically DIS cross sections are separated into leptonic and hadronic tensors where the hadronic tensor can be written in terms of parton distribution functions (PDFs).  These PDFs codify the low-energy QCD structure of the hadron~\cite{Collins2011}.  One would like to compare the experimental extractions of these PDFs with theoretical predictions, which must of necessity be calculated non-perturbatively.

One non-perturbative method of calculating hadronic structure is lattice QCD, whereby one discretizes Euclidean space-time onto a finite hypercube such that cross sections can be calculated numerically from first principles (see for example ~\cite{Gattringer:2010zz}).  However, the PDFs are defined using light-front coordinates which translate poorly to the lattice.  It is possible to calculate moments of the PDFs from the lattice, as in~\cite{Hagler:2007xi,Bali:2012av,Braun:2008ur,Guagnelli:2004ga,Detmold:2003rq,Detmold:2001dv}.  These have not resulted in enough moments or precise enough moments to extract accurate PDFs though some notable progress has been made in this direction~\cite{Chambers:2017dov}.  

A recent idea, developed by X. Ji and collaborators, involves calculating an alternative distribution that reduces to the PDFs in a large-momentum limit~\cite{Ji:2013dva,Xiong:2013bka,Ji:2013fga,Ji:2014gla,Ji:2015jwa,Ji:2015qla,Chen:2016fxx,Ji:2016zfr}.  These new effective-theory PDFs, or quasi PDFs, are time-independent and thus can be calculated directly on the lattice.  Ideally one can then match the quasidistribution functions to the real distribution functions; thereby providing a prediction to test against experiment.  To match the lattice results to the experimental results, one has to understand the analytic behavior of the lattice calculation and of the continuum calculation such that radiative corrections between two different regularization schemes can be reconciled.  A flurry of exploratory papers has been published on lattice calculations of quasidistributions with discussion of how to match their results to the PDFs~\cite{Ma:2014jla,Alexandrou:2015rja,Alexandrou:2016jqi,Lin:2014zya,Lin:2016qia,Chen:2016utp,Ishikawa:2016znu,Monahan:2016bvm,Zhang:2017bzy}.  However, some details of the matching still require further consideration.

Infinities on the lattice are of course regulated by the lattice spacing and one has to match this scheme to the continuum as it is observed by experiment.  We show how to do this explicitly below, using the Wilson action in lattice pertubration theory (LPT) for definiteness.  However, the larger question turns out not to be how to connect LPT to the continuum, but how to connect Minkowski and Euclidean results in the quasidistribution context.

It has been supposed that perturbative Euclidean QCD is equivalent to perturbative Minkowskian QCD, in a given context.  In this paper, we explore this supposition and the subtleties that arise from it.  We find that there appear to be differences in the infrared (IR) region.

We demonstrate these differences by considering one radiative correction, the vertex correction, to the quark quasidistribution and comparing the result to the known Minkowski analog.  We further show how these differences arise and discuss why we should expect them generally in the present context.  In the final discussion, we consider possible resolutions to the Euclidean vs Minkowskian discrepancy.  In particular, we discuss a recent preprint~\cite{Briceno:2017cpo} provided by Brice$\tilde{\text{n}}$o and collaborators which places light on features of the quark quasidistribution as it is defined and calculated on the lattice versus the same as it is calculated below in standard momentum space lattice perturbation theory.

But first, some background is necessary.  That will come in the next section.  Section~\ref{sec:vertex} will contain a presentation of the LPT and continuum one-loop Euclidean space corrections and a discussion of the results is offered in Sec.~\ref{sec:discussion}.

\section{Background}                    \label{sec:background}

The parton distribution function (PDF), or specifically, the quark distribution of a proton is denoted $q(x,\mu^2,P^+)$, where $x$ is the momentum fraction of the quark relative to the proton's momentum $P$, and $\mu^2$ (or $\Lambda^2$) is some normalization scale (or momentum cutoff). It is defined using light-cone coordinates~\cite{Brodsky:2000sk} $\xi^- = (t - z)/\sqrt{2}$ such that

\begin{align} \label{eq:quarkDistributionLC}
    &q(x,\mu^2,P^+)         \nonumber\\
    &\quad = \int_{-\infty}^\infty \frac{d \xi^{-}}{4 \pi} e^{- i x \xi^{-} P^{+}} \braket{P|\bar{\psi}(\xi^{-}) \gamma^{+} W(\xi^{-}) \psi(0) |P}
\end{align}
The gauge link is $W(\xi^-) = \text{exp}\left(-i g \int_0^{\xi^-} d \eta^- A^+ (\eta^-) \right)$, and $A$ is the gluon field.

In the large $P^z$ limit~\cite{Weinberg:1966jm}, X. Ji~\cite{Ji:2013dva} showed that the quark distribution can be given as a purely space-like integral with corrections of orders of $\mathcal{O} ( \Lambda^2/(P^z)^2 )$ and $\mathcal{O} (M^2/(P^z)^2)$, such that

\begin{align} \label{eq:quarkDistributionLCexp}
    q(x,\Lambda^2,P^z) &= \int_{-\infty}^\infty \frac{d z}{4 \pi} e^{i x z P^{z}} \braket{P|\bar{\psi}(z) \gamma^{z} W(z) \psi(0) |P} \nonumber \\ 
        &\quad + \mathcal{O} (\Lambda^2/(P^z)^2 , M^2/(P^z)^2).
\end{align}
Here the gauge link is 
$W(z) = \text{exp}\left(i g \int_0^z dz' A^z (z') \right)$, $\Lambda$ is a transverse momentum regulator, and $M$ is the proton mass.  Ji calls the integral above the quark quasidistribution, denoted $\tilde{q}$, and $\tilde{q} \rightarrow q$ as $P^z \rightarrow \infty$.  A convenient feature of the quasidistribution is the lack of time dependence.  In particular, this allows for one to use lattice QCD to calculate the distribution nonperturbatively.  Then, as long as $P^z$ is taken large enough (\textit{i.e.}, larger than the proton mass $M$ and larger than the inverse lattice spacing $a^{-1}$, the explicit momentum-cutoff regulator on the lattice), one can use this quasidistribution as an accurate representation of the physical quark distribution.  However, to connect the quark distribution and quasidistribution precisely, one needs to consider and compare the regularization schemes that are used in obtaining quark distributions from experimental data and in obtaining the quasidstributions from the lattice.   

\begin{figure}[t!]
\begin{center}
    \includegraphics[width = 84 mm]{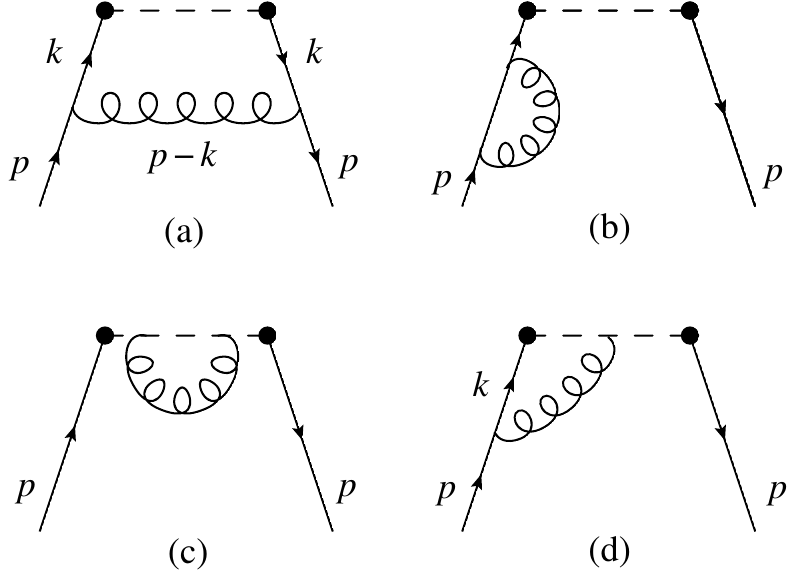}
\caption{The one-loop perturbative corrections to the quark distributions or quasidistributions.  Working in light-front (for the distributions) or axial gauge (for the quasidistributions), diagrams (c) and (d) do not contribute.  $p = (p_0,0,0,p_z)$ is the quark momentum.}
\label{fig:vertices}
\end{center}
\end{figure}

One loop perturbative corrections to the distribution functions are illustrated in Fig~\ref{fig:vertices}.  In suitable gauges, light-front gauge for the PDFs and axial gauge for the quasidistributions, diagrams (c) and (d) of this Figure are absent (see \cite{Konetschny:1977xm} for details when working in axial gauge).  One can then calculate the wave function renormalization $Z_F$ (and $\tilde Z_F$ for the quasidistribution) at order $\alpha_s$ and the vertex corrections $q^{(1)}$ (and $\tilde q^{(1)}$) also at order $\alpha_s$, and relate the distributions by
\begin{align}       \label{eq:quarkDistCorrections}
&q(x,\Lambda) = \tilde q(x) \left( 1 + Z_F (\Lambda) - \tilde Z_F (\Lambda) \right) \nonumber \\
    &\quad + \int_x^1 \frac{dy}{y} \left( q^{(1)}(x/y,\Lambda) -\tilde q^{(1)}(x/y,\Lambda) \right)
        q(y,\Lambda) + \mathcal{O}(\alpha_s^2)     .
\end{align}

For reference, the one-loop vertex correction to the PDF for a quark connected to another quark, at leading order in $m^2/p_z^2$ (where $m$ and $p_z$ denote the quark mass and momentum respectively) is
\begin{align} \label{eq:vertexCorrectionMinkPhys}
&q^{(1)}(x,\Lambda) = \frac{ \alpha_s C_F }{ 2\pi }          \nonumber\\
    &\hskip 3 em \times \bigg\{ \frac{ 1+x^2 }{ 1-x } \ln\frac{ \Lambda^2 }{ (1-x)^2 m^2 }
      - \frac{ 2x }{ 1-x } - \delta_\epsilon (1-x)  \bigg\}   ,
\end{align}
when regulated by a cutoff on the transverse momentum~\cite{Xiong:2013bka}, with $\delta_\epsilon = 0$.  For dimensional regularization, replace $\Lambda$ by a renormalization scale $\mu$ and let $\delta_\epsilon =1$.  Here $C_F = 4/3$ is the standard color factor~\cite{Altarelli1977298}.  The result is non-zero only for $0<x<1$.   The corresponding quasidistribution is non-zero also for $x>1$ and $x<0$.  We will only quote the result~\cite{Xiong:2013bka}, again for reference, for $0<x<1$, calculated in Minkowski space at lowest order in $m^2/p_z^2$,
\begin{align} \label{eq:vertexCorrectionMink}
&\tilde{q}^{(1)}_{\text{M}}(x,\Lambda,p^z) =    \frac{\alpha_s C_F}{2\pi}     \nonumber\\
&\hskip 1.4 em \times \bigg\{ \frac{1 + x^2}{1-x} \ln \frac{4x p_z^2}{(1-x)m^2} 
        - \frac{4x}{1-x} + 1 + \frac{\Lambda}{(1-x)^2 p^z} \bigg\}    .
\end{align}

While the ultraviolet divergences are different, note that the infrared divergence for $m\to 0$ is the same in the two expressions (equations \eqref{eq:vertexCorrectionMinkPhys} and \eqref{eq:vertexCorrectionMink}).  For a review on the logarithmic divergence and how it arises when the momenta of the virtual quark and gluon in Fig.~\ref{fig:vertices} are collinear, see \cite{Amati:1978by,Politzer:1977yc}.

It is important to recall the goal of this program: to compare $\tilde{q}$ calculated via Lattice QCD to $q$ extracted from experiment.  An underlying assumption is that effects from the lattice being discretized are relatively negligible and that effects from the lattice being in Euclidean space are similarly negligible.  

For small enough lattice spacing, the former at first appears to be a reasonable assumption.  One can show that corrections from the lattice spacing only appear at orders of $a m$ or $a p_z$.  However, to match the quasidistribution to the real distribution, we require from \eqref{eq:quarkDistributionLCexp} that $P^z \gg \Lambda$.  On the lattice $\Lambda \sim 1/a$ and thus $a p_z$ is not necessarily a small quantity.  Consequences of this have been discussed in~\cite{Radyushkin:2016hsy,Radyushkin:2017gjd,Radyushkin:2017ffo} and we will only minimally consider them further here.

Additionally, a second assumption should be studied.  We will find that the radiative corrections calculated on Euclidean space have different behavior in the infrared region compared to the same calculated in Minkowski space.  This points to subtleties in using Euclidean lattice perturbation theory techniques to analyze the lattice itself.

We study these two assumptions by considering a particular example, the vertex correction Fig.~\ref{fig:vertices}(a) calculated via LPT whereby one discretizes space-time with the Wilson action from which all the required propagators and vertex rules can be derived.  For a review of lattice perturbation theory see~\cite{Capitani:2002mp,CAPITANI2001313,CAPITANI2001183}.

\section{Vertex Correction}         \label{sec:vertex}

\subsection{Preliminary: continuum Euclidean integrals}
\label{sec:euclid}

We will calculate the vertex correction to the quark quasidistribution, Fig~\ref{fig:vertices}(a), using axial gauge and working in Euclidean space.  As a preliminary, we first calculate in the continuum, and then calculate the leading terms of a small lattice-spacing, $a$, expansion in lattice perturbation theory using the Wilson action.  One also needs the wave-function renormalization at one-loop, Fig.~\ref{fig:vertices}(b), but this can be quickly obtained via quark number conservation~\cite{Xiong:2013bka},
\begin{align}
\int\limits^{\infty}_{-\infty} dx \, \tilde q(x) = \int\limits^1_0 dx \, q(x) = 1   \,,
\end{align}
or (for example) $\tilde Z_F = - \int dx \, \tilde q(x)$.

In the Euclidean space continuum, the vertex correction is
\begin{align}
&\tilde q^{(1)}_{\text{E}}(x,\Lambda,p_z) = \frac{- i g_0^2 C_F}{4\pi} \int \frac{d^3k}{(2\pi)^3}  \nonumber\\
&\hskip 3 em    \times
    \frac{ \bar u(p) \gamma_\mu ( -i \slashed{k} + m ) \gamma_z ( -i \slashed{k} + m ) \gamma_\nu u(p)\, d_{\mu\nu}(p-k) }
         { (k^2+m^2 )^2 (p-k)^2 }   ,
\end{align}
where
\begin{align}
d_{\mu\nu}(q) = \delta_{\mu\nu} - \frac{ q_\mu n_\nu + n_\mu q_\nu }{ n\cdot q} + \frac{ q_\mu q_\nu }{ (n\cdot q)^2 }  ,
\end{align}
with $n$ being a unit vector in the $z$-direction.  The integration measure $d^3k = dk_1 dk_2 dk_4$, and $k_3 = k_z = x p_z$.

Upon working out the numerator, the result can be written as simple factors times five Euclidean-space integrals, which are
\begin{align}
\label{eq:integrals}
I_1 &= \int \frac{d^3k}{(2\pi)^3} \frac{1}{(k^2+m^2) (p-k)^2} = \frac{\ln{2}}{ 8 \pi p_z },     \nonumber\\
I_2 &= \int \frac{d^3k}{(2\pi)^3} \frac{1}{(k^2+m^2)^2} = \frac{1}{ 8 \pi x p_z },     \nonumber\\
I_3 &= \int \frac{d^3k}{(2\pi)^3} \frac{1}{(k^2+m^2)^2 (p-k)^2} =  -\frac{1}{ 32 \pi p_z^3 }\frac{1-x}{x},     \nonumber\\
I_4 &= \int \frac{d^3k}{(2\pi)^3} \frac{p\cdot k}{(k^2+m^2)(p-k)^2} = -\frac{( 1-x ) p_z}{ 8\pi } ,  \nonumber\\
I_5 &= \int \frac{d^3k}{(2\pi)^3} \frac{1}{(p-k)^2} = \frac{1}{ 4 \pi }\left( \Lambda -  p_z \right),  \nonumber\\
\end{align}
where $m$ is a quark mass and $\Lambda$ is again a transverse momentum cutoff larger than $p_z$.  We have, where possible, neglected the quark mass, and placed the quark on-shell in Euclidean space such that $p^2=-m^2$. This last identification is not crucial, as commented upon in Sec.~\ref{sec:discussion}.  Neglecting the quark mass is reasonable for finite $x$, \textit{e.g.,} $x > m/p_z$ (this condition is also required for the Minkowski space result as usually quoted \eqref{eq:vertexCorrectionMink}).

One should note that in Euclidean space the denominators of each integrand above cannot, in general, be combined with the standard Feynman trick because the denominators may change sign in the region of integration.  Instead one evaluates these integrals by first integrating over $k_4$ via contour integration and then integrating over the remaining transverse momentum $k_\perp^2$.

Of these integrals, the first is for the present discussion the most interesting.  The analog of the $I_1$ integral in Minkowski case gives the $\ln(m^2)$ terms:
\begin{align}
\label{eq:1M}
I_{1M} &\equiv -i \int \frac{d^3k}{(2\pi)^3} \frac{1}{(k^2-m^2+i\epsilon) \left( (p-k)^2 + i\epsilon \right)}
        \nonumber\\
    &= \frac{1}{ 8 \pi p^z } \ln\frac{ 4 x (p^z)^2 }{ (1-x) m^2 } ,
\end{align}
for $0<x<1$ with $d^3k = dk^0 dk^1 dk^2$.  We will consider in Section~\ref{sec:poles} the mechanism which causes the results for $I_1$ and $I_{1M}$ to differ.

The Euclidean results for the quasidistribution are,
\begin{align}
\label{eq:quasieuclid}
\tilde{q}^{(1)}_{\text{E}} 
&= \frac{\alpha_s C_F}{2 \pi} \left\{
    \begin{array}{ll}
        \frac{1+x^2}{1-x} \ln 2  - \frac{1}{(1-x)^2} + 2 + \frac{\Lambda}{p_z (1-x)^2}, &0<x<2, \\ \\
        \frac{2x}{1-x} \ln \frac{x-1}{x} - 1 +\frac{\Lambda}{p_z (1-x)^2}, &x<0, \\ \\
        \frac{2x}{1-x} \ln \frac{x}{x-1} - 1 +\frac{\Lambda}{p_z (1-x)^2}, &x>2, \\
    \end{array}
    \right.
\end{align}
with terms of $\mathcal O(m^2/p_z^2)$ not written.

The UV divergent terms are the same as in the Minkowski evaluation~\eqref{eq:vertexCorrectionMink}, but there are no IR divergences for $m\to 0$.

\subsection{Minkowski and Euclidean integrals}
\label{sec:poles}

The lack of an infrared divergence observed in \eqref{eq:quasieuclid} within the region $0<x<1$ points to a fundamental difference between the Euclidean and Minkowskian loop integrals.  Here we analyze how this difference arises by considering the analogue of the Euclidean $I_1$ integral~\eqref{eq:integrals} in Minkowski space~\eqref{eq:1M} which contains the logarithmic infrared divergence.  

\begin{figure}
    \begin{centering}
    \includegraphics[width = 82 mm]{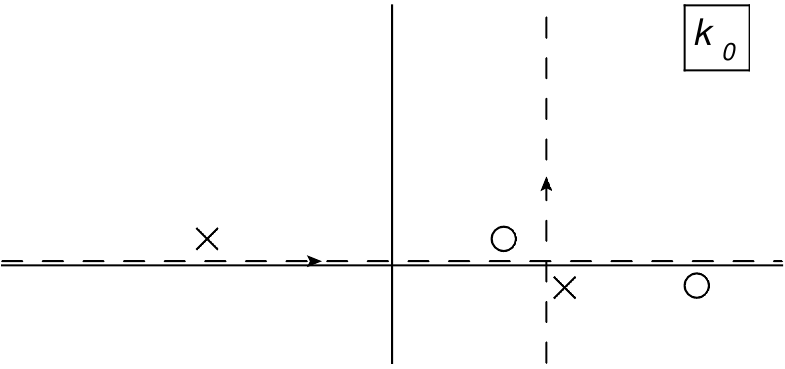}
    \end{centering}
\caption{Poles in the complex $k^0$ plane.  Crosses represent poles from the quark propagator and open circles represent poles from the gluon propagator.}
\label{fig:wick}
\end{figure}

Our strategy is to attempt to convert this Minkowski space integral back into its Euclidean space brother.  We will find that we are forced to cross a pole in the complex $k_4$ plane.

\begin{align}
    I_{1M} = \frac{-i}{(2\pi)^3} \int \frac{dk^0 dk^1 dk^2}{ (k^2 - m^2 + i\epsilon )( (k-p)^2 + i\epsilon) } ,
\end{align}
with $k^z = x p^z$.  There are four poles in $k^0$, two from the quark propagator,
\begin{align}
    k^0 = \pm \sqrt{k_\perp^2 + x^2 p_z^2 + m^2} \mp i\epsilon     ,
\end{align}
and two from the gluon propagator,
\begin{align}
\label{eq:gluonpole}
    k^0 = \sqrt{p_z^2 + m^2} \pm \sqrt{k_\perp^2 + (1-x)^2 p_z^2} \mp i\epsilon     .
\end{align}
The locations of the four poles in the complex $k^0$ plane are shown in Fig.~\ref{fig:wick}, using crosses for the quark poles and open circles for the gluon poles.


The gluon pole in the lower half plane (LHP) is always to the right of the quark pole in the LHP, and the gluon pole in the upper half plane (UHP) is always to the left of the quark pole in the LHP.  The reader may demonstrate the first of these statements.  To demonstrate the second, notice that nonzero $k_\perp$ pushes the UHP gluon pole farther to the left and the LHP quark poles farther to the right, so it suffices to consider the situation for $k_\perp = 0$.  Then for large $p^z$, finite $x$, and small $m$, the UHP gluon pole is at 
\begin{align}
    \left. k^0 \right|_{\rm gluon} = p^z + \frac{m^2}{ 2 p^z } - (1-x) p^z + i \epsilon = x p^z 
        + \frac{m^2}{ 2 p^z } + i \epsilon     ,
\end{align}
while the LHP quark pole is at,
\begin{align}
    \left. k^0 \right|_{\rm quark} = x p^z + \frac{m^2}{ 2 x p^z } - i \epsilon .
\end{align}
Since for $0<x<1$,
\begin{align}
    \frac{m^2}{ 2 x p^z } > \frac{m^2}{ 2 p^z }     ,
\end{align}
the statement, and the relative location of the poles in Fig.~\ref{fig:wick}, are verified.  (In Fig.~\ref{fig:wick}, the UHP gluon pole could be farther to the left than illustrated, including possibly to the left of the imaginary axis.)

Since the integral is convergent, and given the locations of the poles,  there is always a region on the real axis, say including a point $k^0 = E_x$, about which we can Wick rotate the $k^0$ integration line.  The $90^\circ$ rotated integration line is also shown on Fig.~\ref{fig:wick}.  If we define
\begin{align}
    k_4 = i k^0
\end{align}
(and $p_4 = i p^0$), we can write the Minkowski integral as
\begin{align}
    &I_{1M} =  \frac{1}{(2\pi)^3}\int d^2k_\perp \int_{-\infty + i E_x}^{\infty + i E_x} dk_4   \quad \times     \nonumber\\
    &\frac{1}{ ( k_4^2 + k_\perp^2 + x^2 p_z^2 + m^2 )( (k_4-p_4)^2 + k_\perp^2 + (1-x)^2 p_z^2 ) } ,
\end{align}


\begin{figure}[t]
    \begin{centering}
    \includegraphics[width = 83 mm]{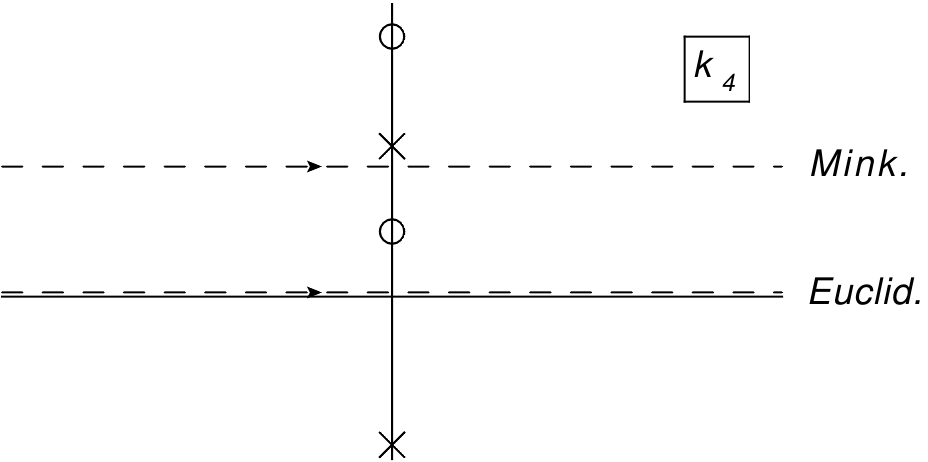}
    \end{centering}
\caption{Integration lines and poles of the integrand shown for the complex $k_4$ plane.  A gluon propagator pole lies between the two integration paths unless $k_\perp$ is large enough.}
\label{fig:k4plane}
\end{figure}


This is the same as the Euclidean integral ($I_{1}$ of \eqref{eq:integrals}), except that in the Euclidean case the $dk_4$ integral runs along the real line.  The two integration lines, and the poles of the integrand, are shown in the complex $k_4$ plan in Fig.~\ref{fig:k4plane}.

The two integration lines give the same results if the integrand has no poles between them.  But this is not always the case. A gluon pole lies between the two integration lines when $k_\perp$ is small, specifically when (from Eq.~(\ref{eq:gluonpole})),
\begin{align}
    k_\perp^2 < p_z^2 - (1-x)^2 p_z^2 + m^2 = x(2-x) p_z^2 + m^2    .
\end{align}

Hence we cannot move the integration line, and the two integrals, Minkowski and Euclidean, are not the same.  This argument can be extended for all finite loop-momentum integrals though the location of the poles may differ.

The initial integration paths are not arbitrary. In Euclidean LPT the momenta range over real numbers and correspond to integrals on the real line in the Euclidean continuum, and for Minkowski space the energy integration must be along the real line.  The configuration of the integration contour or line after a Wick rotation is dictated by the movement of the poles with changing kinematics, and another example of this phenomenon may be seen in~\cite{Hart:2006ij}.

There is in Sec.~\ref{sec:discussion} further discussion of the difference between the Euclidean and Minkowski evaluations.


\subsection{Integrals on the lattice}

Using the Wilson action with Wilson parameter $r$~\cite{Capitani:2002mp},  the one-loop vertex correction (Fig.~\ref{fig:vertices}(a)) is,
\begin{align} \label{eq:vertexCorrection}
\tilde q^{(1)}_L(x,p_z,a) =  & \frac{-i}{2} g^2_o C_F \mathlarger{\int\limits_{-\pi/a}^{\pi/a}} \!\!\! \frac{\,d^4k}{(2\pi)^4} \delta (k_z - x p_z) G_{\rho \lambda}(p-k)  \nonumber \\
&\times  \bar u(p) V_\rho(p,k) S(k) \gamma_z S(k) V_\lambda(k,p)  u(p)  \,.
\end{align}
where 
\begin{align}
S(k) &= \frac{1}{  i \sum_\mu \gamma_\mu \sin a k_\mu + a m_0 + 2r \sum_\mu \sin^2 (a k_\mu /2) }  \label{eq:quarkProp} \\
G_{\rho \lambda}(l) &= \frac{1}{ \hat l_\mu^2}
    \Big( \delta_{\rho\lambda} - \frac{ \hat l_\rho \hat n_\lambda + \hat n_\rho \hat l_\lambda }{ \hat n \cdot \hat l}
    + \frac{ \hat l_\rho \hat l_\lambda }{ (\hat n \cdot \hat l)^2 } 
    \Big) \label{eq:gluonProp} \\
 \hat l _\rho &= \frac{2}{a} \sin (a l_\rho / 2) ,  \nonumber\\
 \hat n _\rho &= n_\mu \cos (a l_\rho / 2) ,  \nonumber\\
V_\rho(p,k) &= i\gamma_\rho \cos(a(p+k)_\rho /2 ) + r \sin( a (p+k)_\rho /2)  ,  \label{eq:vertex}
\end{align}

Again, we have chosen to work in axial gauge such that the gauge link $W$ in \eqref{eq:quarkDistributionLC} is unity and Fig.~\ref{fig:vertices}(a) is the only diagram that contributes to $\tilde{q}^{(1)}$ at first order in $\alpha_s$.  One can check that the analogs of the first four integrals given in the continuum case, Eq.~\eqref{eq:integrals}, remain finite and in the limit of lattice spacing going to zero, have the same values as previously given.

We now consider the ultraviolet divergent integral, the analog of $I_5$ in Eq.~\eqref{eq:integrals}, arising from the longitudinal $(p-k)_\rho (p-k)_\lambda$ piece of the axial gauge gluon propagator Eq.~\eqref{eq:gluonProp},
\begin{align}\label{eq:divInt}
I_{5L} &= \mathlarger{\int\limits_{-\pi/a}^{\pi/a}} \!\!\!\!\! \frac{d^4 k}{(2\pi)^3} \delta (k_z - x p_z) 
\frac{a^2/4}{ \sum_{\mu=1}^4 \sin^2 \left( a(p_\mu-k_\mu)/2 \right) }   \nonumber\\
&=  \frac{1}{4a} \int\limits_{-\pi}^\pi \frac{ d^3K }{ (2\pi)^3 }
    \Big[ \sin^2 \Big( \frac{ ap_4-K_4 }{ 2 } \Big) + \sin^2 \Big( \frac{K_1}{2} \Big) \nonumber\\
&\hskip 4 em    + \sin^2 \Big( \frac{K_2}{2} \Big)
        + \sin^2 \Big( \frac{a(1-x)p_z}{2} \Big)   \Big]^{-1}       .
\end{align}
We can take $a \rightarrow 0$ within the integral in the second form, and thus obtain the divergent term.  Additionally evaluating the integral numerically for finite lattice spacings and analyzing the results, we find
\begin{align}\label{eq:divIntCont}
I_{5L} &= \left\{
    \begin{array}{ll}
        \frac{1}{4 \pi} \left( \frac{n}{a} -p_z \right) + \mathcal{O}(a p_z) + \mathcal{O} (m^2/p_z^2), \quad 0<x<2, \\ \\
        \frac{1}{4 \pi} \left( \frac{n}{a} -(1-x)p_z  \right)+ \mathcal{O}(a p_z) + \mathcal{O} (m^2/p_z^2), \quad x<0, \\ \\
        \frac{1}{4 \pi} \left( \frac{n}{a} -(1-x)p_z  \right)+ \mathcal{O}(a p_z) + \mathcal{O} (m^2/p_z^2), \quad x>2, \\
    \end{array}
    \right.
\end{align}
with $n=3.17591$, a numerical factor representing the slope of the ultraviolet divergence.  One may think of $n/a$ as being the radius of a circle which approximates the transverse momentum square.  That is, $n$ is the numerical necessity that results from using polar coordinates to integrate \eqref{eq:divInt} analytically.  

At order $a^{-1}$ and $a^0$, the Wilson parameter $r$ does not affect the perturbative result. Further, we find no evidence of a $\log{a}$ term.  We notice the simple correspondence from $I_5$ in Eq.~\eqref{eq:integrals}
\begin{align}
    \Lambda \leftrightarrow \frac{n}{a}     .
\end{align}
Furthermore, we note the existence of discretization corrections in the corners of the momentum cube that go like $a p_z$.  Knowing that $\Lambda \sim 1/a$, these corrections aren't necessary small in the region where one can expand the PDFs into quasidstributions \eqref{eq:quarkDistributionLCexp}.  That is, to calculate the quasidistribtion via LPT in hopes of obtaining the physical distribution, we are forced to choose $a p_z > n > 1$.

At the lowest orders in $a$ and $m/p_z$, we find $\tilde{q}^{(1)}$ is
\begin{align}\label{eq:vertexCorrectionLat}
\scriptstyle \tilde{q}^{(1)}_{\text{Lat.}}
&= \scriptstyle \frac{\alpha_s C_F}{2 \pi} \left\{
    \begin{array}{ll}
        \frac{1+x^2}{1-x} \ln 2  - \frac{1}{(1-x)^2} + 2 + \frac{n}{a p_z (1-x)^2}, & \scriptstyle 0<x<2, \\ \\
        \frac{2x}{1-x} \ln \frac{x-1}{x} - 1 +\frac{n}{a p_z (1-x)^2}, & \scriptstyle x<0, \\ \\
        \frac{2x}{1-x} \ln \frac{x}{x-1} - 1 +\frac{n}{a p_z (1-x)^2}, & \scriptstyle x>2. \\
    \end{array}
    \right.
\end{align}

\begin{figure}[t!]
\includegraphics[width=\columnwidth]{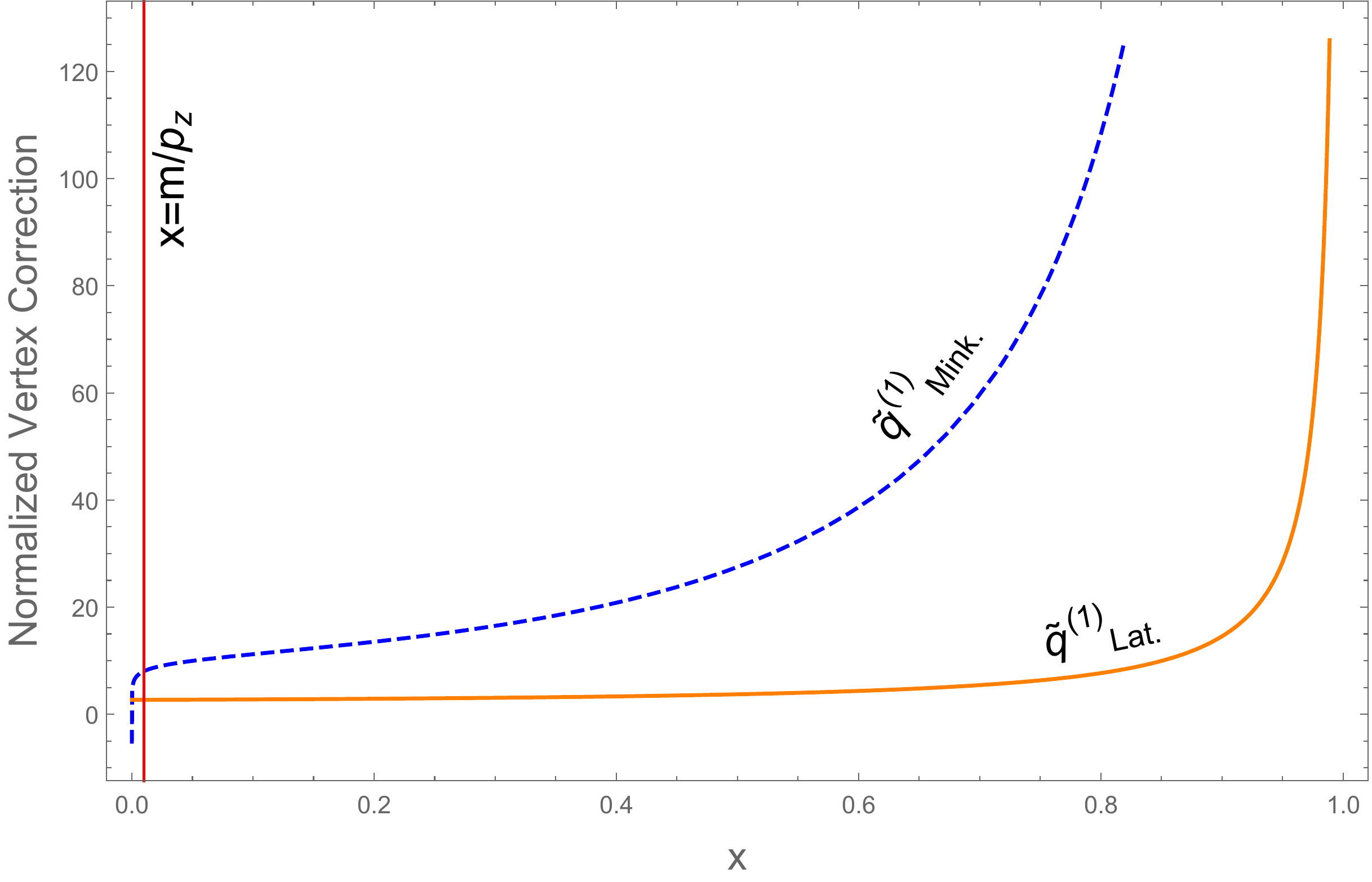}
\caption{The vertex correction to the quark quasidistribution, normalized by $\alpha_s C_F / (2\pi)$, calculated in Minkowski space (blue dashed line) versus the same calculated via LPT in Euclidean space (orange solid line) to leading order in $m^2/p_z^2$ for momentum fraction $x$ between $0$ and $1$.  Here $p_z$ is $2$ GeV, and the quark mass $m$ is $0.02 \text{\ GeV}$.  $\Lambda \leftrightarrow n/a$ is $2 \text{\ GeV}$.  When $\Lambda < p_z$ the sign of the $x \rightarrow 1$ pole flips for the LPT (orange solid line) result while the Minkoski result remains qualitatively unaffected.  The red bar denotes the region where $x = m/p_z$.  To the left of this line, the expansion in $m^2/p_z^2$ is not valid.}
\label{fig:qMinkVSqLat}
\end{figure}

Absent from \eqref{eq:vertexCorrectionLat} is the infrared divergence as $m \rightarrow 0$ featured in the Minkowski result \eqref{eq:vertexCorrectionMink}.  This results from the qualitative difference between the Euclidean and Minkowski calculations previously discussed, and the numerical difference for some selected parameters is seen in Fig.~\ref{fig:qMinkVSqLat}.


Also, as a small note, we reiterate that both the LPT result \eqref{eq:vertexCorrectionLat} and the continuum result \eqref{eq:vertexCorrectionMink} are given at order $m^2/p_z^2$.  The expansion in $m/p_z$ is not valid in the region when the momentum fraction $x \ll m/p_z$.  This is seen in the fact that there is a discontinuity between the quoted $x<0$ and $x>0$ results.  The vertex correction for general $m$ can be calculated and smooths out this region.  There is a logarithmic divergence in $m$ for the lattice result at $x = 0$.  However, this feature is nonexistent for finite $x$ and thus does not change the qualitative differences between the Euclidean LPT and the Minkowskian continuum noted above.

\section{Discussion}                        \label{sec:discussion}

We found that non-trivial differences occur between loop corrections calculated in Euclidean space (in LPT) and the same calculated in Minkowski space (on the continuum).  These differences are not manifested purely as small corrections resulting from the discretization of space-time.  Rather, the Euclidean space behavior of loop-integrals is qualitatively different from their Minkowskian counterparts.

Specifically, the infrared divergence (for quark mass going to zero) that is manifest in the Minkowski space evaluations is absent in the Euclidean evaluations.  Previously calculated perturbative corrections in Minkowski space for the regular PDFs and for the quasidistributions had shown the same IR divergences.  This was considered a positive feature.  The low momentum parts of the perturbative calculation might be unreliable, but if they were the same one could argue that the IR region did not require correction if one used a quasidistribution to calculate the PDFs.  The differences in the UV region could be corrected using the perturbative results, since they are reliable at high momenta.

The lack of an infrared divergence in the LPT vertex correction \eqref{eq:vertexCorrectionLat} is disquieting.  At a minimum, it means that more thought is needed about how one can calculate the quasidistribution on the lattice, necessarily in Euclidean space, and then correct just the UV parts of the result to obtain the usual PDFs.

In the main text, in Sec.~\ref{sec:poles}, we gave a pole-based argument regarding the differences in the Euclidean and Minkowski results.  One can also give a more physical argument.  In the Minkowski case, the infrared divergence comes from a near collinear configuration that occurs when the quark and gluon in the loop are moving parallel and both nearly on-shell.  That they can never be absolutely parallel in 4D is only because of the mass of the quark.  In the Euclidean case, the parallel momentum situation is not reached.  In Sec.~\ref{sec:vertex}, we put the external quarks on-shell even in the Euclidean case, and found no IR divergence.  In a general numerical calculation in Euclidean space, the quark outside the loop has all momenta real, and in such a case it is still clearer that there will be no collinear singularity and no chance to have a mass singularity.   The conclusions about the infrared behavior of the Euclidean case are not dependent on the specific choice of external four-momentum.

It may be that further thought will reconcile the Euclidean LPT and Minkowski continuum evaluations.  In the full hadron case, the incoming and outgoing quarks seen in Fig.~\ref{fig:vertices} are intermediate states and their momenta are integrated over.  It was proved some time ago (see for example~\cite{Ellis:1978sf,Ellis:1978ty}) in Minkowski space that the leading power contributions come from the case where the quark is on-shell.  In a numerical calculation in Euclidean space, the quark with all momentum components real cannot be on-shell and knowing what replaces the former on-shell contributions may be relevant to connecting the Euclidean and Minkowski calculations.  

Motivated by the arguments noted above, work in this direction has been considered by R. Brice$\tilde{\text{n}}$o and collaborators~\cite{Briceno:2017cpo}.  There it is argued that one may overcome the present observations in the infrared region by consideration of a calculation more directly mimicking that which is done on the lattice, with its integration over all spatial points and its large separation of temporal points.  Such a consideration leads to the continuum Euclidean results found here plus an additional pole contribution compensating for the difference between the Euclidean and Minkowski results.  In particular, the lattice analogue of the quark quasidistribution does not correlate incoming and outgoing quark states of equal and definite energy as one would do in standard perturbation theory or in LPT. 

In considering perturbative corrections in the UV region, one can start from the standard Euclidean Wilson (or other lattice) action and Euclidean LPT but applied to correlation functions as defined by the lattice itself, rather than corrections calculated for momentum eigenstates, as is commonly done in Minkowski space, and which are the corrections calculated for example in~\cite{Xiong:2013bka}.  In this method, the on-shell contribution is not necessarily the only leading power contribution and one has to carefully analyze additional leading-order contributions that may arise from the integration over the incoming and outgoing momenta.  However, at least in the present case, the UV divergence found in the LPT vertex correction \eqref{eq:vertexCorrectionLat} remains the same.

\begin{acknowledgments}

We thank Raul Brice$\tilde{\text{n}}$o, Christopher Monahan, Kostas Orginos, Anatoly Radyushkin, Jianwei Qiu, and Savvas Zafeiropoulos for helpful conversations.  We thank the National Science Foundation for support under Grant PHY-1516509, and MF thanks the U.S. Department of Energy for support under contract DE-AC05-06OR23177, under which Jefferson Science Associates manages Jefferson Lab.

\end{acknowledgments}

\bibliography{main.bib}

\end{document}